%% file: igaed.tex
\pdfoutput=1
\documentclass[conference]{IEEEtran}
\IEEEoverridecommandlockouts

\usepackage{graphicx}
\usepackage{color}
\usepackage{pgf, tikz, pgfplots}
\usepackage{siunitx}
\pgfplotsset{compat=1.17} 
\usepackage{amssymb} 
\usepackage{bm}
\usepackage{acronym}
\usepackage{tikz}
\usepackage{mathrsfs}
\usepackage{enumerate}

\usepackage[utf8]{inputenc} 
\usepackage[T1]{fontenc}
\usepackage{url}
\usepackage{ifthen}
\usepackage{cite}
\usepackage[cmex10]{amsmath} %

\input{jonathan_definitions}
\begin{document} 
\title{Improved Generalized Automorphism\\ Belief Propagation Decoding\\
\thanks{This work has received funding from the 
German Federal Ministry of Education and Research (BMBF) within the project Open6GHub (grant agreement 16KISK010) and the European Research Council (ERC) under the European Union’s Horizon 2020 research and innovation programme (grant agreement No. 101001899).}
}

\author{\IEEEauthorblockN{Jonathan Mandelbaum, Sisi Miao, Nils Albert Schwendemann, Holger Jäkel, and Laurent Schmalen}
\IEEEauthorblockA{Communications Engineering Lab, Karlsruhe Institute of Technology (KIT), 76131 Karlsruhe, Germany\\
\texttt{jonathan.mandelbaum@kit.edu}}
}
\maketitle

\begin{abstract}
With the increasing demands on future \mbox{wireless} \mbox{systems}, new design objectives become eminent. 
Low\--density parity\--check codes together with belief propagation (BP)  \mbox{decoding} have outstanding performance
for large block lengths. Yet, for future wireless systems, good decoding performance for short block lengths is mandatory, a regime in which 
BP decoding typically shows a significant gap to maximum likelihood \mbox{decoding}. Automorphism ensemble decoding (AED) is known to reduce this gap effectively and, in addition, enables an easy trade-off \mbox{between} latency, throughput, and complexity.
Recently, \mbox{generalized} AED (GAED) was proposed to increase the set of feasible \mbox{automorphisms} suitable for ensemble decoding.
By construction, GAED requires a preprocessing step within its constituent paths that results in information loss and potentially limits the gains of GAED.
In this work, we show that the preprocessing step can be merged with the Tanner graph of BP decoding, thereby improving the performance of the constituent paths.
Finally, we show that the improvement of the individual paths also enhances the overall performance of the ensemble.
\end{abstract}

\section{Introduction}
With the ongoing rollout of 5G, research activities shift their focus onto the analysis of enabling technologies for the next-generation 6G wireless systems \cite{8869705}. The rising demand for flexibility to adapt to different use cases, e.g., \ac{URLLC}, 
also impacts the channel coding scheme. This shift yields new expected design objectives, e.g., low latency and reduced decoding complexity \cite{10274812,10104534}. Thereby, unified coding is a candidate for becoming an enabling technology for a future 6G channel coding scheme\cite{10274812}. Herein, unified coding refers to a single code family that performs well in the different specified scenarios by, e.g., employing different decoders \cite{10274812}. 
In particular, \ac{LDPC} codes, first introduced by Gallager \cite{Gallager_LDPC_diss} together with a low-complexity iterative decoding scheme often denoted as \ac{BP} decoding, have outstanding decoding performance for large block lengths \cite{MCT08}. However, for future use cases, e.g., \ac{URLLC} scenarios, short block length codes are unavoidable, a regime in which the \ac{BP} decoding performance of \ac{LDPC} shows a significant gap to the performance of \ac{ML} decoding.
Yet, \ac{BP} decoding provides benefits, e.g., low latency and simple hardware implementation.
Therefore, one step towards a potential unified coding scheme is to improve the performance of \ac{BP} based decoding schemes in the short block length regime.

Ensemble decoding schemes such as, e.g., \ac{MBBP}\cite{MBBP2,MBBP3_withLeaking}, \ac{AED} \cite{AED_RMcodes,stuttgart_ldpc_aed}, \ac{GAED}\cite{mandelbaum2023generalized}, and \ac{EED}\cite{mandelbaum2024endomorphisms}, can enhance the performance of BP decoding for short block length codes and provide flexibility in the trade-off of complexity, throughput, and 
latency.
They exploit the fact that the successful decoding of the received word at the output of a \ac{BMSC}
with BP decoding is determined by the realization of the noise, rather than the respective codeword \cite[Lemma 4.90]{MCT08}.
Automorphism-based ensemble decoding algorithms, i.e., \ac{AED} and \ac{GAED}, perform parallel decoding operations in parallel paths each relying on an 
alternative noise representation.\footnote{In addition, \ac{AED} and \ac{GAED} allow for different decoding algorithms in their respective paths, e.g.,
 using BP decoding based on different \acp{PCM} and, hence, incorporating ideas of \ac{MBBP} \cite{MBBP2}. Note that in each path, a PCM with redundant rows may be used.}

The most prominent definition of the automorphism group of channel codes stems from \cite{MacWilliamsSloane} where the authors define the automorphism group of a binary code to consist of all
permutations that are self-mappings of the code. Recently, in \cite{mandelbaum2023generalized}, we proposed a more general definition of the automorphism group defining it as the set of all linear bijective self-mappings. Furthermore, we introduced the concept of \ac{GAED} demonstrating that generalized automorphisms can be used for decoding  \cite{mandelbaum2023generalized}.
Hereby, the paths of a \ac{GAED} require a preprocessing step which results in an information loss and reduces the decoding gains achieved by the ensemble scheme.

In this work, we improve \ac{GAED} by reducing the losses due to the preprocessing. To this end,
we first demonstrate that the preprocessing required for \ac{GAED} can be merged with the Tanner graph of the code, resulting in a larger Tanner graph which is then used for decoding. We name this algorithm \ac{iGAED}.
Herein, ``improved'' refers to two possible benefits achieved by this construction: On the one hand, decoding performance with respect to error rate is improved by avoiding information loss during preprocessing, particularly in the high SNR regime.
On the other hand, implicitly realizing mapping (\ref{eq:boxplus}) within the Tanner graph may be advantageous for hardware implementation.

\section{Preliminaries}
\subsection{Linear Block Codes}
A binary linear block code $\mathcal{C}(n,k)$ consists of $2^k$ distinct vectors that form a $k$-dimensional subspace of the vector space $\mathbb{F}_2^n$, where the parameters $n\in \mathbb{N}$ and $k\in \mathbb{N}$ denote block length and information length, respectively. For simplicity, we omit the parameters $(n,k)$ when obvious from the context. Note that vectors are column vectors.
Linear block codes can be expressed as the null space of their parity\--check matrix (PCM) $\bm{H}\in \mathbb{F}_2^{m\times n}$\cite{MacWilliamsSloane}, i.e.,
$$\mathcal{C}\left(n,k\right)=\left\{\bm{x}\in \mathbb{F}_2^n:\bm{H} \bm{x} = \bm{0}\right\}=\mathrm{Null}(\bm{H}).$$

 \Ac{BP} decoding is an iterative message passing algorithm on the Tanner graph of the code \cite{MCT08}. A Tanner graph is a bipartite graph with two disjoint sets of vertices: the \acp{VN} $\mathfrak{V}$ corresponding to the code bits $x_i$, and, thus, the columns of $\bm{H}$, and the \acp{CN} $\mathfrak{C}$ corresponding to the parity checks and, thus, to rows of $\bm{H}$. A VN ${\mathsf{v}_i\in\mathfrak{V}}$ is connected to a CN $\mathsf{c}_j\in\mathfrak{C}$ if the corresponding entry $H_{j,i}=1$. 
Messages are iteratively propagated along the edges and updated in the nodes of the graph \cite{MCT08}.
 A linear code can be expressed by different Tanner graphs.
 The performance of BP decoding dominantly depends on the degrees of the nodes and the short cycles within the Tanner graph and, hence, the performance typically differs when decoding on the different Tanner graph representations of a code \cite{MCT08}.

\subsection{Automorphism Group}
In \cite{MacWilliamsSloane}, the automorphism group of a binary linear code is defined as
\begin{equation*}
\mathrm{Aut}(\mathcal{C}):=    
\left\{ \pi: \mathcal{C}\to\mathcal{C}, \bm{x}\mapsto\pi(\bm{x}): \pi\in S_n   \right\},
\end{equation*}
with ${\pi(\bm{x})=\begin{pmatrix}
x_{\pi(1)},&\ldots &,x_{\pi(n)}
\end{pmatrix}}^\mathsf{T}$ and $S_n$ being the symmetric group\cite{MacWilliamsSloane}. Therefore, it consists of all index permutations that map codewords onto codewords.
In \cite{mandelbaum2023generalized}, a different view on the automorphism group of codes is introduced, using the more general notion prominent in linear algebra.
To this end, the generalized automorphism group of a binary linear code 
$\mathcal{C}$ is defined as
\begin{equation*}
\mathrm{GAut}(\mathcal{C}):=    
\left\{ \tau: \mathcal{C}\rightarrow\mathcal{C}: \text{ $\tau$ linear, $\tau$ bijective} \right\}.
\end{equation*}

The (permutation) automorphism group $\mathrm{Aut}(\mathcal{C})$ is a subgroup of the generalized automorphism group $\mathrm{GAut}(\mathcal{C})$~\cite{mandelbaum2023generalized}. Furthermore, note that a linear mapping ${\tau:\mathcal{C}\rightarrow\mathcal{C}}$ can be expressed by a transformation matrix ${\bm{T}\in \mathbb{F}_2^{n\times n}}$,
i.e., \begin{equation}\label{eq:linear_mapping}
\bm{x}_\tau:=
\tau(\bm{x})=\bm{T}\bm{x}\in \mathcal{C}.
\end{equation}
Since $\tau\in\mathrm{GAut}(\mathcal{C})$ is only assumed to be bijective on $\mathcal{C}$ and not on $\mathbb{F}_2^n$, its transformation matrix $\bm{T}$ is not necessarily non-singular. However, for the sake of simplicity,
we only consider cases in which $\bm{T}^{-1}$ exists, as in \cite{mandelbaum2023generalized}.

\subsection{Generalized Automorphism Ensemble Decoding}

\begin{figure}[t]
    \centering
    \begin{tikzpicture}
    \tikzset{myblock/.style={rectangle, draw, thin, minimum width=0.6cm, minimum height=0.6cm},font=\footnotesize,align=center}%
\tikzset{mywideblock/.style={rectangle, draw, thin, minimum width=1cm, minimum height=0.6cm},font=\footnotesize,align=center}%
        \node (input) at (-1.5,-0.5) {$\bm{y}$};    
        \draw[fill] (-1.02,-0.5) circle (1pt);
        \node (pi1) [rectangle, draw, thin, minimum width=1.6cm, minimum height=0.6cm] at (0.2,0.3) {Pre$(\bm{y}|\bm{T}_1$)};
        \node (pik) [rectangle, draw, thin, minimum width=1.6cm, minimum height=0.6cm] at (0.2,-1.3) {Pre$(\bm{y}|\bm{T}_K$)};
        
        \node (debox_1) [myblock, right=0.75cm of pi1] {BP($\bm{H}$)};
        \node (debox_K) [myblock, right=0.75cm of pik] {BP($\bm{H}$)};
        
        \node (pi_inv_1) [rectangle, draw, thin, minimum width=1cm,  minimum height=0.6cm,right=0.75cm of debox_1]  {$\bm{T}_1^{-1}$};
        \node (pi_inv_k) [rectangle, draw, thin, minimum width=1cm,  minimum height=0.6cm,right=0.75cm of debox_K]  {$\bm{T}_K^{-1}$};
        
        \node (ml_1) [right=0.7cm of pi_inv_1] {};
        \node (ml_K) [right=0.7cm of pi_inv_k] {};
        
        \node (ml_box) [rectangle, draw, thin, minimum width=0.6cm, minimum height=2.2cm] at (5.61,-0.5)  {\rotatebox{90}{ML-in-the-list}};
        \node (output) [right=0.3cm of ml_box] {$\widehat{\bm{x}}$};
        
        \node (dots_box) [below=0cm of pi1]  {$\vdots$};
        \node (dots_box) [below=0cm of debox_1]  {$\vdots$};
        \node (dots_pi_inv_box) [below=0cm of pi_inv_1] {$\vdots$};
        \myline{input}{pi1};
        \myline{input}{pik};
        
        \draw [-latex] (pi1) -- (debox_1) node[draw=none,fill=none,font=\scriptsize,midway,above] {${\bm{L}}_{\tau_1}$};
        \draw [-latex] (pik) -- (debox_K) node[draw=none,fill=none,font=\scriptsize,midway,above] {${\bm{L}}_{\tau_K}$};
        
        \draw [-latex] (debox_1) -- (pi_inv_1) node[draw=none,fill=none,font=\scriptsize,midway,above] {$\widehat{{\bm{x}}}_{\tau_1}$};
        \draw [-latex] (debox_K) -- (pi_inv_k) node[draw=none,fill=none,font=\scriptsize,midway,above] {$\widehat{{\bm{x}}}_{\tau_K}$};
        
        \draw [-latex] (pi_inv_1.east) -- (ml_1) node[draw=none,fill=none,font=\scriptsize,midway,above] {$\widehat{\bm{x}}_1$};
        \draw [-latex] (pi_inv_k) -- (ml_K)  node[draw=none,fill=none,font=\scriptsize,midway,above] {$\widehat{\bm{x}}_K$};
        
        \draw [-latex] (ml_box) -- (output);	
    
    \end{tikzpicture}
    \caption{Block diagram of \ac{GAED} introduced by \cite{mandelbaum2023generalized} in which $K$ different automorphisms with transformation matrices ${\bm{T}_\ell }$ are used. We assume that each path performs BP decoding based on the PCM $\bm{H}$.}\label{figure:gaed}
    \vspace{-0.5cm}
\end{figure}
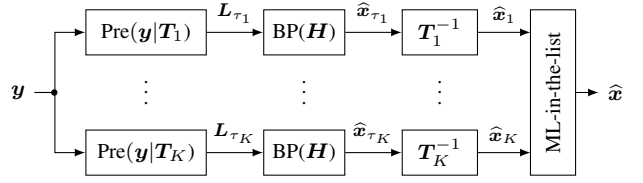

    \begin{figure*}[ht]
        \centering
        \includegraphics[scale=1]{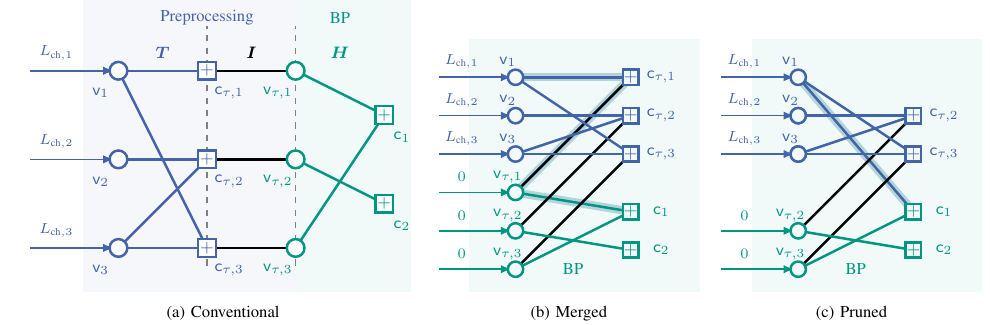}
        \caption{Graphical illustration of the transformation of the conventional
        path from GAED in (a) via the merged Tanner graph in (b) to the pruned Tanner graph in (c).}
        \label{fig:illustration}
        \vspace*{-1.5em}
    \end{figure*}

\ac{AED} as a generalization of \ac{MBBP} decoding was first introduced in \cite{AED_RMcodes} and further generalized to \ac{GAED} in \cite{mandelbaum2023generalized}.
Figure~\ref{figure:gaed} depicts a block diagram of \ac{GAED} based on BP decoding which we will briefly recapitulate.

Consider the transmission of a binary codeword $\bm{x}\in \mathcal{C}\subset \mathbb{F}_2^n$ over a \ac{BMSC}.
At the channel output, $\bm{y}\in\mathcal{Y}^n$ is observed (with $\mathcal{Y}$ being the output alphabet) and then mapped onto the bit-wise LLR vector
$\bm{L}:=\left(L(y_j|x_j)\right)_{j=1}^n\in \mathbb{R}^n$.
Then, if ${\bm{T}_1,\ldots,\bm{T}_K}$ are the non-singular transformation matrices of $K$ generalized automorphisms used for \ac{GAED}, the ensemble decoding consists of $K$ parallel paths each including a preprocessing block, a BP decoding block, and an inverse mapping block.

In path $\ell\in \left\{1,\ldots, K\right\}$, the $j$th element, $j \in \{1,\ldots,n\}$, of bit-wise LLR vector $\bm{L}$ is preprocessed according to\cite{mandelbaum2023generalized}:
\begin{equation}\label{eq:boxplus}
\left({\bm{L}}_{\tau_\ell}\right)_j%
:=\left(\text{Pre}(\bm{y}|\bm{T}_\ell)\right)_j:= %
\mathboxplus_{
\begin{subarray}{c}
i=1,\\
T_{\ell,j,i}=1 \end{subarray}}^{n} \!\!\!\! L(y_i|x_i),%
\end{equation}
with the box-plus operator $\mathboxplus$ introduced in \cite{485714} 
to imitate the $\mathbb{F}_2$~summation in (\ref{eq:linear_mapping}) in the LLR domain. 
Note that variables with subscript $\tau$, e.g., ${\bm{L}}_{\tau_\ell}$, denote values after preprocessing as introduced in (\ref{eq:linear_mapping}). 
If $\bm{T}_\ell$ is the transformation matrix of a permutation automorphism $\tau_\ell\in \mPermAut$ then the preprocessing corresponds to the permutation of LLRs known from \ac{AED} \cite{mandelbaum2023generalized}. 

Next, the preprocessed LLR vector ${\bm{L}}_{\tau_\ell}$ is decoded by a BP decoder, resulting in an estimate
$\widehat{{{\bm{x}}}}_{\tau_\ell}$.
If BP decoding converges to a codeword, i.e., if $\widehat{{{\bm{x}}}}_{\tau_\ell}\in \mathcal{C}$, then $\widehat{{{\bm{x}}}}_{\tau_\ell}$ constitutes an estimate of 
the codeword after the preprocessing, i.e.,
$\widehat{{{\bm{x}}}}_{\tau_\ell}=\widehat{\bm{T}_\ell  \bm{x}}_\ell$.
Hence, to compensate for the effect of preprocessing, $\widehat{{{\bm{x}}}}_{\tau_\ell}$ has to be post-processed with the inverse automorphism $\bm{T}_\ell^{-1}$ to obtain an estimate of the originally transmitted codeword, i.e., $ {\widehat{\bm{x}}_\ell=\bm{T}_\ell^{-1}\widehat{{\bm{x}}}_{\tau_\ell}}$.

Finally, GAED collects the $K$ estimates of all paths in an \emph{ML-in-the-list} block that outputs the final estimate $\widehat{\bm{x}}$ according to the \emph{ML-in-the-list} rule described in \cite{AED_RMcodes}.

In \cite{mandelbaum2023generalized}, the impact of the \emph{weight over permutation} $${\Delta\left(\bm{T}\right)=\sum_{i,j} T_{i,j} - n}$$
is analyzed. It is demonstrated that larger weight over permutations leads to an increasing information loss due to the preprocessing that results in a degradation of the performance of that path. Yet, as shown in \cite{mandelbaum2023generalized}, GAED still can exploit the additional diversity of the degraded parallel paths to improve the ensemble's performance. Thus, GAED is a promising candidate to enhance decoding performance, particularly for short block length codes. However, open questions remain. First, each path performs its respective preprocessing once and then ignores the additional knowledge about the code structure provided by the automorphism. Second, the decoder in each path has no access to the originally received LLR vector $\bm{L}$ but only to the preprocessed values $\bm{L}_{\tau_\ell}$.
Thus, in the following section, we propose an alternative perspective aiming to avoid explicit preprocessing by merging it with BP decoding and, thus, iteratively
and jointly exploiting the knowledge of the automorphism and the unprocessed LLR vector during decoding.
Hence, we expect an improved performance of the ensemble by avoiding information loss in the preprocessing of the respective paths.

\section{Improved Generalized Automorphism Belief Propagation Decoding }

In this section, we improve the individual paths of GAED by avoiding explicit preprocessing as described in (\ref{eq:boxplus}) and implicitly performing the according mapping within \ac{BP} of a suitably extended Tanner graph.

\subsection{Preprocessing as Extended Tanner Graph}
To this end, we realize the preprocessing as part of a larger Tanner graph 
consisting of $2n$ \acp{VN} forming the set $\mathfrak{V}'$ and ${n+m}$ \acp{CN} forming the set $\mathfrak{C}'$, respectively. The extended Tanner graph simultaneously represents the structure of the PCM of the code and describes the effects of the preprocessing due to generalized automorphism $\bm{T}$. Figure \ref{fig:illustration} illustrates the transformation process of the graph.

As illustrated in Fig.~\ref{fig:illustration}(a), the extended Tanner graph is defined to consist of $2n$ \acp{VN} ${\mathfrak{V}'=\mathfrak{V}\!\cup\!\mathfrak{V}_\tau}$ with ${\mathfrak{V}:=\left\{\mathsf{v}_{i}: i=1,\ldots, n\right\}}$ and $\mathfrak{V}_\tau:=\left\{\mathsf{v}_{\tau,i}:i=1,\ldots, n\right\}$. %
Herein, $\mathsf{v}_{i}$ is denoting the $i$th %
code bit of the estimated transmitted codeword 
and $\mathsf{v}_{\tau,i}$ is denoting the {$i$th} position 
of the estimated preprocessed codeword 
after applying the generalized automorphism $\bm{T}$, respectively.

The set of check nodes $\mathfrak{C}'$ of the extended Tanner graph is defined by ${\mathfrak{C}'=\mathfrak{C}\!\cup\!\mathfrak{C}_\tau}$ with $\mathfrak{C}$ denoting the original check nodes, described by PCM $\bm{H}$, and $\mathfrak{C}_\tau$ denoting additional check nodes representing the preprocessing. 
Note that, according to (\ref{eq:boxplus}), each row operation of the generalized automorphism $\bm{T}$ can be interpreted as a CN, in total yielding $n$ additional \acp{CN} that form the set $\mathfrak{C}_\tau$.

In order to consider the fact that $\mathsf{v}_{i}$ is contributing to the {$j$th} preprocessed value $\mathsf{v}_{\tau, j}=(\bm{T}\bm{\mathsf{v}})_j$, the following connections are constituted: 
\begin{itemize}
    \item VN $\mathsf{v}_{i}\in \mathfrak{V}$ is connected to (new) CN $\mathsf{c}_{\tau,j}$  if and only if $T_{j,i}\neq0$, i.e., if $\mathsf{v}_{i}$ is participating in $\mathsf{v}_{\tau, j}$

    \item CN $\mathsf{c}_{\tau,j}$ is connected to (new) VN $\mathsf{v}_{\tau,j}\in\mathfrak{V}_\tau$ and not connected to any other $\mathsf{v}_{\tau, \ell}\in\mathfrak{V}_\tau, \ell\neq j$ %
\end{itemize}

Therefore, the Tanner graph associated with the preprocessing, connecting $\mathfrak{V}$ and $\mathfrak{V}_\tau$ by CNs $\mathfrak{C}_\tau$, is considered as the Tanner graph of the PCM
\[
\bm{H}_{\text{pp}}=\begin{pmatrix}
    \bm{T} &\bm{I}_{n\times n}
\end{pmatrix}\in \mathbb{F}_2^{n\times 2n}.
\]
of a \ac{LDGM} code.
Now, the GAED preprocessing step is realized by a %
message passing of the channel information from the \acp{VN} $\mathfrak{V}$ in a check node update in $\mathfrak{C}_\tau$ and a \ac{VN} update in \acp{VN} $\mathfrak{V}_{\tau}$. 

Next, in GAED, as illustrated in the right-hand side of Fig.~\ref{fig:illustration}(a), BP decoding is performed on the Tanner graph associated with the original PCM $\bm{H}$ using the preprocessed values in $\mathfrak{V}_{\tau}$ %
and the \acp{CN} in $\mathfrak{C}$.
In Fig.~\ref{fig:illustration}(a), the unprocessed channel information is unavailable during BP decoding.

We propose to transform the graph into the structure depicted in
Fig.~\ref{fig:illustration}(b). Therein, we stack the $n$ \acp{VN} in $\mathfrak{V}_{\tau}$ below the $n$ \acp{VN} in $\mathfrak{V}$ and the $m$ CNs in $\mathfrak{C}_\tau$ above the $n$ CNs in $\mathfrak{C}$, respectively, while maintaining the edges.
This yields a larger Tanner graph, consisting of $2n$ VNs $\mathfrak{V}'$ and $n+m$ CNs $\mathfrak{C}'$, associated with a PCM
\begin{equation}\label{eq:H:iGAED}
\bm{H}_{\text{iGAED}}=\begin{pmatrix}
    \bm{T} &\bm{I}_{n\times n}\\
    \bm{0}_{m\times n} &\bm{H}\\
\end{pmatrix}\in \mathbb{F}_2^{(n+m)\times 2n}.    
\end{equation}

Hereby, the first $n$ VNs are connected to the LLR vector $\bm{L}$, i.e., to $\mathfrak{V}$ and possess unprocessed channel information.  The remaining $n$ \acp{VN} account for the auxiliary checks in $\mathfrak{V}_\tau$, representing the preprocessing.
No a-priori information is known about the VNs in $\mathfrak{V}_\tau$ 
which, therefore, are considered as erased.\footnote{Another possibility is to use the preprocessing from GAED to obtain information about the \acp{VN} from $\mathfrak{V}_\tau$. However, our simulation results showed equal performances with respect to error rates for both setups. Hence, we erase the additional bits to avoid the explicit calculation of the preprocessing step.} 
Note that the merging process can also be interpreted as a different decoding schedule based on the graph depicted in Fig.~\ref{fig:illustration}(a) where all VNs and CNs are updated in parallel, respectively.

\subsection{Pruning of the Extended Tanner Graph}
To simplify the structure of the Tanner graph, note that a transformation matrix $\bm{T}$ with weight over permutation ${\Delta(\bm{T})<n}$ possesses at least one row with a single non-zero entry, resulting in a CN $\mathsf{c}_{\tau, j}$ of degree $2$ in the Tanner graph defined by $\bm{H}_{\text{pp}}$. %
Assume that $\mathsf{c}_{\tau, j}$ is connected to $\mathsf{v}_{i}\in\mathfrak{V}$ and $\mathsf{v}_{\tau, \ell}\in\mathfrak{V}_\tau$.
Then, similar to \cite{8437581}, %
the extended Tanner graph can be simplified by merging the \ac{VN} $\mathsf{v}_{\tau, \ell}\in\mathfrak{V}_\tau$ into the VN $\mathsf{v}_{i}\in\mathfrak{V}$, rerouting all other connections of $\mathsf{v}_{\tau, \ell}$ to $\mathsf{v}_i$. Note that a VN after merging corresponds to both a code bit of the received codeword $\bm{y}$ as well as a bit within the preprocessed codeword $\bm{y}_{\tau}$. Furthermore, note that the adaptation of $\bm{H}_{\text{iGAED}}$ can be realized by column additions and row/column deletions.
The PCM after pruning is denoted by $\bm{H}_{\text{iGAED,p}}$.

 \begin{figure}[t]
    \centering
    \begin{tikzpicture}	[level 1/.style={sibling distance=20mm},edge from parent/.style={->,draw},>=latex]
    \tikzset{myblock/.style={rectangle, draw, thin, minimum width=0.6cm, minimum height=0.6cm},font=\footnotesize,align=center}%
\tikzset{mywideblock/.style={rectangle, draw, thin, minimum width=1cm, minimum height=0.6cm},font=\footnotesize,align=center}%
		\node (in) at (0,0) {$\bm{y}$};
		\node (prepro) [myblock,right=0.3cm of in] {Embedding};
		\node (dec) [myblock, right=2cm of prepro] {BP$\left(\bm{H}_{\text{iGAED,p}}\right)$};
		\node (out) [right=0.6cm of dec] {$\hat{\bm{x}}$};
		\draw[->] (in) -- (prepro);
		\draw[->](prepro)-- (dec)  node[midway,above]{$\begin{pmatrix}
		    \bm{L}^{\mathsf{T}}\,\,\,\bm{0}_{1\times n}^{\mathsf{T}}
		\end{pmatrix}^{\mathsf{T}}$} ;
		\draw[->] (dec) -- (out);
	\end{tikzpicture}%
    \vspace*{-2mm}%
    \caption{A path of iGAED  using the pruned matrix $\bm{H}_{\text{iGAED,p}}$. The embedding maps the channel output to an extended LLR vector of dimension $2n$ by appending zeros.} \label{figure:aut_preprocessing}
 \end{figure}
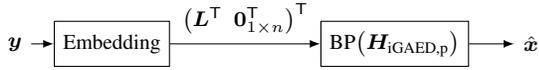
 
A single path of iGAED is depicted in Fig.~\ref{figure:aut_preprocessing}. Thereby, the channel output is mapped to its LLR, extended to length $2n$ as described beforehand, and BP decoding is performed using the Tanner graph associated with $\bm{H}_{\text{iGAED}}$ or its pruned version $\bm{H}_{\text{iGAED,p}}$, where the VNs associated with an actual code bit possess channel information and the others are 
initialized by zero LLRs.

Note that \ac{iGAED} involves a higher number of CNs and VNs, resulting in a slight increase in complexity compared to \ac{GAED}.

\emph{Example:} 
The transformation matrix\footnote{Note that the transformation $\bm{T}$ used in this example and in Fig.~\ref{fig:illustration} does not constitute an actual automorphism, but is chosen for simplicity of illustration.} and the PCM of Fig.~\ref{fig:illustration} are given by
\[
\bm{T}
=
\begin{pmatrix}
    1 & 0 & 0 \\
    0 & 1 & 1\\
    1 & 0 & 1
\end{pmatrix}
, \quad
\bm{H}
=
\begin{pmatrix}
    1 & 0 & 1\\
    0 & 1 & 0
\end{pmatrix},
\]
respectively, resulting in 
\[
\bm{H}_{\text{iGAED}}
=
\begin{pmatrix}
    1 & 0 & 0 & 1 & 0 & 0\\
    0 & 1 & 1 & 0 & 1 & 0\\
    1 & 0 & 1 & 0 & 0 & 1 \\
    0 & 0 & 0 & 1 & 0 & 1\\
    0 & 0 & 0 & 0 & 1 & 0    
\end{pmatrix}.
\]

\ac{CN} $\mathsf{c}_{\tau, 1}$ has only two neighbors $\mathsf{v}_1\in\mathfrak{V}$ and $\mathsf{v}_{\tau,1}\in\mathfrak{V}_\tau$. Thus, \acp{VN} $\mathsf{v}_1$ and $\mathsf{v}_{\tau,1}$ can merged into VN $\mathsf{v}_1\in\mathfrak{V}$, see Fig.~\ref{fig:illustration}(c). Edge $(\mathsf{v}_{\tau, 1}, \mathsf{c}_1)$ is rerouted as $(\mathsf{v}_{1}, \mathsf{c}_1)$. 

Merging $\mathsf{v}_{\tau, 1}$ and $\mathsf{v}_{1}$ is achieved
by adding the column associated to $\mathsf{v}_{\tau, 1}$ onto the column associated to $\mathsf{v}_{1}$ and deleting the row associated to $\mathsf{c}_{\tau, 1}$ yielding the pruned matrix

$$\bm{H}_{\text{iGAED,p}}
=
\begin{pmatrix}
    0 & 1 & 1 & 1 & 0\\
    1 & 0 & 1 & 0 & 1 \\
    1 & 0 & 0 & 0 & 1\\
    0 & 0 & 0 & 1 & 0    
\end{pmatrix}.%
$$

\section{Results}

\begin{table}
        \centering
        \caption{Code parameters \cite{mandelbaum2023generalized}}\label{table:constructed_codes}
        \vspace*{-2mm}
        \begin{tabular}{cccccc}
                \toprule
                \textbf{Code} &     $\mathcal{C}_2$ & $\mathcal{C}_{\mathrm{CCSDS}}$& $\mathcal{C}_3$  &   $\mathcal{C}_{\mathrm{BCH}}$  \\
                \midrule
                                    $n$&         $32$&  $32$  & $63$    & $63$\\
                                $k$&            $16$&   $16$  & $45$    & $45$\\
                        $d_\text{min}$ &         $5$ &   4  &    $5$     & $7$\\
                        $\Delta(\bm{T})$&        10&    --&      5&--\\
        \,\,\,\,\,\,\,$\Delta(\bm{T}^{-1})$&     13&    --&      6&--\\
                     \,\,$\Delta(\bm{T}^2)$&      19&  --&      11&--\\
            \,\,\,\,\,\,$\Delta(\left(\bm{T}^{2}\right)^{-1})$&    24&    --&      12&--\\
                \bottomrule
        \end{tabular}
\end{table}

To demonstrate the validity of our approach, we consider two codes $\mathcal{C}_2(32,16)$ and $\mathcal{C}_3(63,45)$ from \cite{mandelbaum2023generalized} with weight over permutation ${\Delta(\bm{T})=10}$ and ${\Delta(\bm{T})=5}$, respectively. In \cite{mandelbaum2023generalized}, codes with associated automorphisms are designed based on the Frobenius normal form.
For comparison, we include two reference codes, namely $\mathcal{C}_{\text{CCSDS}}(32,16)$ from  \cite{channelcodes} and the \ac{BCH} code $\mathcal{C}_{\text{BCH}}(63,45)$.
We analyze the \ac{FER} of different decoding algorithms
by performing Monte\--Carlo simulations. All BP decodings use normalized min-sum decoding with normalization factor $\frac{3}{4}$ and flooding schedule \cite{msa:normalizationconstant}.\footnote{Note that sum-product algorithm yields almost identical performances.}

When designing automorphisms, we exploit that the automorphism group is closed under multiplication and inversion of elements. Thus, given $\bm{T}$, we obtain additional automorphisms $\bm{T}^{-1}$, $\bm{T}^2$, and $\left(\bm{T}^{2}\right)^{-1}$ using basic $\mathbb{F}_2$ matrix operations.
The code parameters and the weights over permutation of the generalized automorphisms are collected in Table~\ref{table:constructed_codes}.
The minimum Hamming distances and approximate performance of \ac{ML} decoding of $\mathcal{C}_2$, $\mathcal{C}_3$, and $\mathcal{C}_{\text{BCH}}$ are taken from \cite{mandelbaum2023generalized}, whereas the minimum Hamming distance of $\mathcal{C}_{\text{CCSDS}}$ is calculated using the algorithm proposed in \cite{search_dmin} and the performance of ML decoding stems from \cite{channelcodes}.

\begin{figure}
    \centering
    \input{figures/ISTC32_16}
    \vspace*{-0.5em}
    \caption{Performance of different decoders for code $\mathcal{C}_2(32,16)$ from \cite{mandelbaum2023generalized}. Additionally, the stand-alone performance of the auxiliary paths of {GAED\--$3$\--BP\--$30$} and {iGAED\--$3$\--BP\--$30$} are depicted, respectively. }
    \label{fig:istc3216}
       \vspace*{-.5em}
\end{figure}
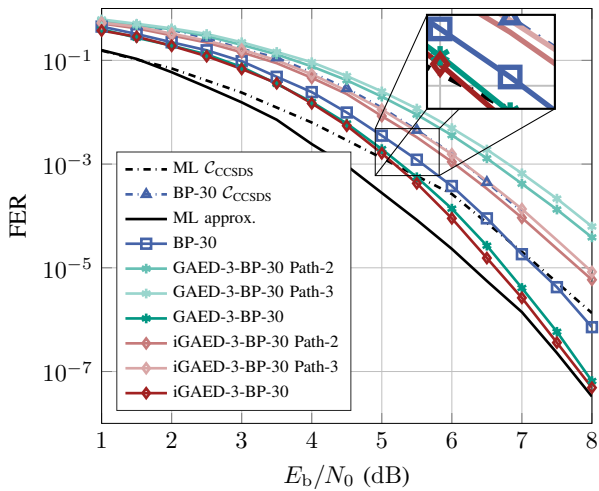
\begin{figure}
    \centering
    \input{figures/ISTC63_45}
    \vspace*{-0.5em}
    \caption{Performance of different decoders for code $\mathcal{C}_3(63,45)$ from \cite{mandelbaum2023generalized}. Additionally, the stand-alone performance of the auxiliary paths of {GAED\--$3$\--BP\--$30$} and {iGAED\--$3$\--BP\--$30$} are depicted, respectively.}
    \label{fig:istc6345}
       \vspace*{-2em}
\end{figure}
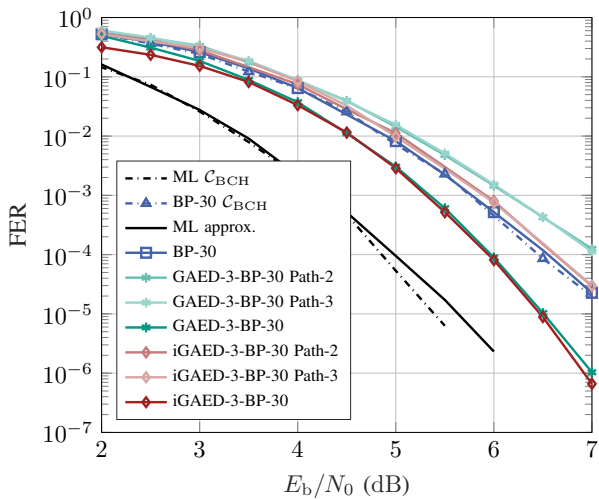

Similar to the notation used in \cite{mandelbaum2023generalized},  {BP\--$p$} denotes BP decoding using $p$ iterations
and {GAED\--$K$\--BP\--$p$} refers to GAED consisting of $K$ paths comprising BP decoding with $p$ iterations each.
Following this convention, {iGAED\--$K$\--BP\--$p$} corresponds to BP decoding performing $p$ iterations on $K$ merged and pruned graphs associated with $\mathbf{H}_\mathrm{iGAED,p}$ as depicted in Fig.~\ref{figure:aut_preprocessing}.
Sparse PCMs for the codes $\mathcal{C}_2$, $\mathcal{C}_3$, and $\mathcal{C}_{\text{BCH}}$ are obtained 
using the algorithm proposed in \cite{8464884}. They are used for BP decoding in GAED and as input for the 
merging and pruning process.%

All GAED and iGAED use the identity mapping $\bm{T}_1=\bm{I}$ in {$\text{Path-}1$}, corresponding to plain BP that typically possesses the best performance among all individual paths. To express this fact, the supplementary paths are termed auxiliary paths from now on. {GAED\--$3$\--BP\--$p$} and {iGAED\--$3$\--BP\--$p$} additionally employ $\bm{T}_2=\bm{T}$ and $\bm{T}_3=\bm{T}^{-1}$ in their auxiliary paths, whereas {GAED\--$5$\--BP\--$p$} as well as {iGAED\--$5$\--BP\--$p$} additionally use paths relying on $\bm{T}_4=\bm{T}^2$ and $\bm{T}_5=\left(\bm{T}^{2}\right)^{-1}$. 
To address different use cases of a possible 6G standard, we consider a \emph{reliable} scenario operating at a target FER of $10^{-3}$ and an \emph{ultra-reliable} scenario
operating at a target FER of $10^{-5}$.

Figure \ref{fig:istc3216} and Fig.~\ref{fig:istc6345} illustrate the \ac{FER} over $E_{\mathrm{b}}/N_0$ of {GAED\--$3$\--BP\--$30$} and {iGAED\--$3$\--BP\--$30$} for the codes 
$\mathcal{C}_2$ and $\mathcal{C}_3$ compared to the performances of their respective constituent paths. 
Additionally, we depict the \ac{ML} performance of the codes as well as the \ac{ML} and BP\--$30$ performance of the respective reference codes.

For both codes, {GAED\--$3$\--BP\--$30$} and {iGAED\--$3$\--BP\--$30$} significantly improve decoding performance compared to stand-alone BP\--$30$.
For code $\mathcal{C}_2$, we observe a gain of $0.45\,\si{dB}$ and $0.5\,\si{dB}$ at an FER of $10^{-3}$
which increases to $0.5\,\si{dB}$ and $0.6\,\si{dB}$ at an FER of $10^{-5}$, respectively. Therefore, the proposed {iGAED\--$3$\--BP\--$30$} yields an additional gain of $0.05\,\si{dB}$ and $0.1\,\si{dB}$ compared to conventional {GAED\--$3$\--BP\--$30$} in the reliable and ultra-reliable scenario, respectively.

Note that the designed automorphism $\bm{T}$ of code $\mathcal{C}_2$ as well as its inverse $\bm{T}^{-1}$ possess a relatively high weight over permutation in the order of $n/3$, resulting in a notable information loss during preprocessing. Consequently, at an FER of $10^{-3}$, {GAED\--$3$\--BP\--$30$} Path-2 and Path-3 require an additional SNR of $1\,\si{dB}$ and $1.1\,\si{dB}$ compared to BP-30, respectively. 
Considering {iGAED\--$3$\--BP\--$30$} Path-2 and Path-3, the SNR gap of the auxiliary paths to BP\--$30$ is reduced to $0.5\,\si{dB}$ and $0.6\,\si{dB}$, respectively.
Hence, the improvement of {iGAED\--$3$\--BP\--$30$} compared to {GAED\--$3$\--BP\--$30$} can be attributed to the significantly better performance of the auxiliary paths. 
Furthermore, it is interesting to highlight that the performance of {iGAED\--$3$\--BP\--$30$} almost fully closes the gap to the \ac{ML} performance in the high SNR regime.

For code $\mathcal{C}_3$, {GAED\--$3$\--BP\--$30$} and {iGAED\--$3$\--BP\--$30$} show a similar gain of $0.5\,\si{dB}$  compared to BP\--$30$ at an FER of ${10^{-3}}$ increasing to $0.6\,\si{dB}$ at an FER of ${10^{-5}}$. Due to the relatively low weight over permutation in the order of $n/10$ of the employed automorphisms, the SNR degradation of the auxiliary paths of {GAED\--$3$\--BP\--$30$}  only amounts to $0.4\,\si{dB}$, significantly lower than for $\mathcal{C}_2$. 
Still, the proposed path structure improves the performance of the auxiliary paths yielding a respective gain of $0.2\,\si{dB}$ at an FER of {$10^{-3}$} for both iGAED-Path-2 and iGAED-Path-3 compared to GAED-Path-2 and GAED-Path-3.
However, as depicted in Fig.~\ref{fig:istc6345}, due to the relatively small weight over permutation, {iGAED\--$3$\--BP\--$30$} does not yield a clear gain in terms of error rate compared to {GAED\--$3$\--BP\--$30$} for both target FERs.
Yet, it can observed that {iGAED\--$3$\--BP\--$30$} yields an equal performance up to an $E_{\mathrm{b}}/N_0$ of $5.5\,\si{dB}$ and possesses better performance than {GAED\--$3$\--BP\--$30$} for higher SNR.

Next, to demonstrate the flexibility of ensemble decoding schemes such as GAED and iGAED, we increase the number of parallel paths, i.e., we increase the complexity while maintaining the latency. Additionally, we compare ensemble decoding at reduced latency to BP decoding with approximately equal complexity.

\begin{figure}
    \centering
    \input{figures/ISTC32_16_diff_paths}
    \vspace*{-0.5em}
    \caption{Performance of {GAED\--$K$\--BP\--$30$} and {iGAED\--$K$\--BP\--$30$} for code $\mathcal{C}_2(32,16)$ from \cite{mandelbaum2023generalized} for $3$ and $5$ parallel paths, respectively.}
    \label{fig:istc3216diffpath}
       \vspace*{-0.5em}
\end{figure}
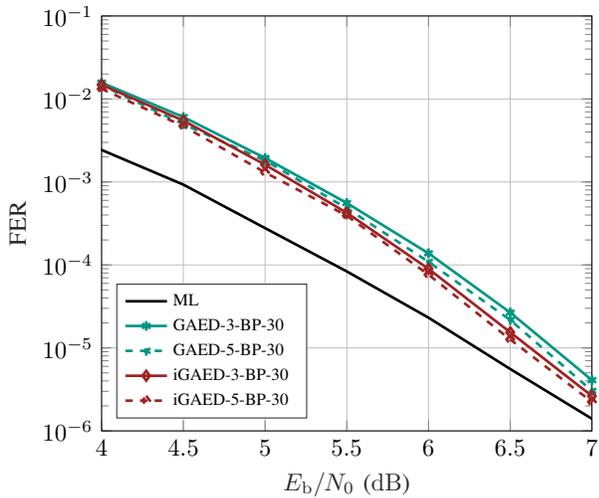
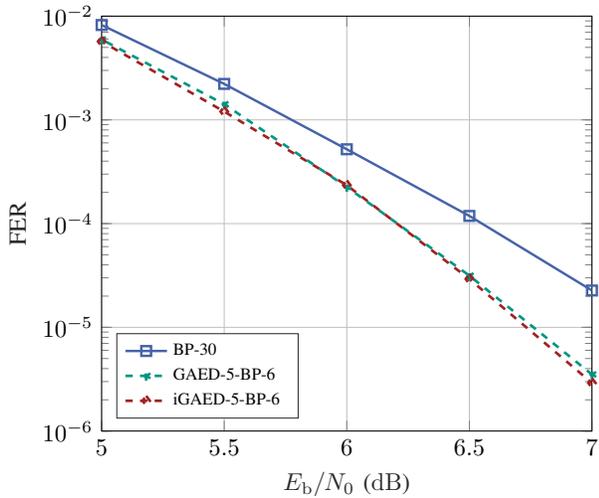
\begin{figure}
    \centering
    \input{figures/ISTC63_45_diff_iter}
    \vspace*{-0.5em}
    \caption{Performance of {GAED\--$5$\--BP\--$6$} and {iGAED\--$5$\--BP\--$6$} for code $\mathcal{C}_3(63,45)$ from \cite{mandelbaum2023generalized} in comparison to BP\--$30$, i.e., at comparable complexity.}
    \label{fig:istc6345diffiter}
    \vspace*{-1.5em}
\end{figure}

To this end, Fig.~\ref{fig:istc3216diffpath} depicts
the \ac{FER} over $E_{\mathrm{b}}/N_0$ of {GAED\--$5$\--BP\--$30$} and {iGAED\--$5$\--BP\--$30$} comparing them to {GAED\--$3$\--BP\--$30$} and {iGAED\--$3$\--BP\--$30$} when decoding $\mathcal{C}_2$.
{GAED\--$5$\--BP\--$30$} yields the lowest FER over the full whole simulated SNR regime. At a constant latency, 
in the ultra-reliable scenario {GAED\--$5$\--BP\--$30$} yields a gain of $0.05\,\si{dB}$, $0.1\,\si{dB}$, and $0.15\,\si{dB}$
compared to {iGAED\--$3$\--BP\--$30$}, {GAED\--$5$\--BP\--$30$} and {GAED\--$3$\--BP\--$30$}, respectively. 
Thereby, it is interesting to highlight that {iGAED\--$3$\--BP\--$30$} outperforms {GAED\--$5$\--BP\--$30$} at both target \acp{FER}.
Therefore, at a constant maximum latency, ensemble decoding schemes can provide the possibility to trade off complexity vs. performance. 

Furthermore, due to the very short block length of the considered codes and the usage of normalized min-sum decoding, their application in a potential ultra-low latency application of 6G is inherent. To further elaborate this use case, we consider parallel decoding using only $6$ iterations of min-sum decoding to reduce the decoding latency. Figure~\ref{fig:istc6345diffiter} depicts the \ac{FER} over $E_{\mathrm{b}}/N_0$ of {GAED\--$5$\--BP\--$6$} and
{iGAED\--$5$\--BP\--$6$} in comparison to BP\--$30$, such that the ensemble decoding schemes have five times lower latency but equal complexity when compared to the stand-alone BP decoding. Thus, two observations are noticeable. First, in particular at the target FERs,
{iGAED\--$5$\--BP\--$6$} tends to yield a slightly lower FER than {GAED\--$5$\--BP\--$6$}. Second, at a constant complexity and reduced latency, {iGAED\--$5$\--BP\--$6$} and {GAED\--$5$\--BP\--$6$} significantly outperform BP\--$30$. At an FER of $10^{-3}$, {iGAED\--$5$\--BP\--$6$} yields a gain of $0.3\,\si{dB}$ compared to BP\--$30$.

\section{Conclusion}

In this work, we showed that the preprocessing of \ac{GAED} can be merged with the Tanner graph associated with BP decoding, yielding an extended Tanner graph which is then used for decoding. This extended graph inherently combines knowledge on the preprocessing by the generalized automorphism with the BP decoding steps. 
Simulations have shown that \ac{iGAED} can improve decoding performance compared to \ac{GAED}, the benefit depending on the weight over permutation.
Additionally, we demonstrated that \ac{iGAED} yields significant improvement compared to \ac{BP} decoding in different use-cases incorporating an \ac{URLLC} scenario.
Moreover, we observed possible close-to-ML performance of \ac{iGAED} in the high SNR regime, showing that a purely BP-based scheme can close the gap to ML decoding in a short block length regime, depending on the code structure.
An interesting open question for future works is the analysis of different decoding schedules on the pruned Tanner graph.

\end{document}

%% file: jonathan_definitions.tex
\newcommand{\myline}[2]{
\path(#1.east) --(#2.west)  coordinate[pos=0.4](mid);
\draw[-latex] (#1.east) -| (mid) |- (#2.west);
}

\usepackage{cite}
\usepackage{amsmath,amssymb,amsfonts}

\usepackage{dsfont}
\usepackage{algorithmic}
\usepackage{graphicx}
\usepackage{textcomp}
\usepackage{xcolor}
\def\BibTeX{{\rm B\kern-.05em{\sc i\kern-.025em b}\kern-.08em
    T\kern-.1667em\lower.7ex\hbox{E}\kern-.125emX}}

    \usepackage{amsfonts}     
\usepackage{graphicx}
\usepackage{color}
\usepackage{pgf, tikz, pgfplots}
\usepackage{siunitx}
\pgfplotsset{compat=1.17} 
\usepackage{amssymb} 
\usepackage{bm}
\usepackage{bbm}
\usepackage{amsthm}
\usepackage{acronym}
\usepackage{tikz}
\usepackage{mathrsfs}
\usepackage{enumerate}
\usetikzlibrary{arrows,calc,fit,matrix,positioning,shapes,shadows,trees,mindmap,tikzmark,arrows.meta,angles,quotes,babel,spy}
\usepackage{pgfplotstable}
\usepgfplotslibrary{groupplots}

\usepackage{booktabs}

\DeclareMathOperator*{\mathboxplus}{\boxplus}

\DeclareMathOperator*{\mPermAut}{\mathrm{Aut}(\mathcal{C})}

\definecolor{kit-green100}{rgb}{0,.59,.51}

\definecolor{kit-green100}{rgb}{0,.59,.51}
\definecolor{kit-green70}{rgb}{.3,.71,.65}
\definecolor{kit-green50}{rgb}{.50,.79,.75}
\definecolor{kit-green30}{rgb}{.69,.87,.85}
\definecolor{kit-green15}{rgb}{.85,.93,.93}
\definecolor{KITgreen}{rgb}{0,.59,.51}

\definecolor{KITpalegreen}{RGB}{130,190,60}
\colorlet{kit-maigreen100}{KITpalegreen}
\colorlet{kit-maigreen70}{KITpalegreen!70}
\colorlet{kit-maigreen50}{KITpalegreen!50}
\colorlet{kit-maigreen30}{KITpalegreen!30}
\colorlet{kit-maigreen15}{KITpalegreen!15}

\definecolor{KITblue}{rgb}{.27,.39,.66}
\definecolor{kit-blue100}{rgb}{.27,.39,.67}
\definecolor{kit-blue70}{rgb}{.49,.57,.76}
\definecolor{kit-blue50}{rgb}{.64,.69,.83}
\definecolor{kit-blue30}{rgb}{.78,.82,.9}
\definecolor{kit-blue15}{rgb}{.89,.91,.95}

\definecolor{KITyellow}{rgb}{.98,.89,0}
\definecolor{kit-yellow100}{cmyk}{0,.05,1,0}
\definecolor{kit-yellow70}{cmyk}{0,.035,.7,0}
\definecolor{kit-yellow50}{cmyk}{0,.025,.5,0}
\definecolor{kit-yellow30}{cmyk}{0,.015,.3,0}
\definecolor{kit-yellow15}{cmyk}{0,.0075,.15,0}

\definecolor{KITorange}{rgb}{.87,.60,.10}
\definecolor{kit-orange100}{cmyk}{0,.45,1,0}
\definecolor{kit-orange70}{cmyk}{0,.315,.7,0}
\definecolor{kit-orange50}{cmyk}{0,.225,.5,0}
\definecolor{kit-orange30}{cmyk}{0,.135,.3,0}
\definecolor{kit-orange15}{cmyk}{0,.0675,.15,0}

\definecolor{KITred}{rgb}{.63,.13,.13}
\definecolor{kit-red100}{cmyk}{.25,1,1,0}
\definecolor{kit-red70}{cmyk}{.175,.7,.7,0}
\definecolor{kit-red50}{cmyk}{.125,.5,.5,0}
\definecolor{kit-red30}{cmyk}{.075,.3,.3,0}
\definecolor{kit-red15}{cmyk}{.0375,.15,.15,0}

\definecolor{KITpurple}{RGB}{160,0,120}
\colorlet{kit-purple100}{KITpurple}
\colorlet{kit-purple70}{KITpurple!70}
\colorlet{kit-purple50}{KITpurple!50}
\colorlet{kit-purple30}{KITpurple!30}
\colorlet{kit-purple15}{KITpurple!15}

\definecolor{KITcyanblue}{RGB}{80,170,230}
\colorlet{kit-cyanblue100}{KITcyanblue}
\colorlet{kit-cyanblue70}{KITcyanblue!70}
\colorlet{kit-cyanblue50}{KITcyanblue!50}
\colorlet{kit-cyanblue30}{KITcyanblue!30}
\colorlet{kit-cyanblue15}{KITcyanblue!15}

\acrodef{AED}{automorphism ensemble decoding}
\acrodef{GAED}{generalized \ac{AED}}
\acrodef{iGAED}{improved \ac{GAED}}
\acrodef{EED}{endomorphism ensemble decoding}
\acrodef{MBBP}{multiple bases belief propagation}

\acrodef{BCH}{Bose--Chaudhuri--Hocquenghem}
\acrodef{FER}{frame error rate}
\acrodef{BSC}{binary symmetric channel}
\acrodef{DE}{density evolution}
\acrodef{LDPC}{low-density parity-check}
\acrodef{LDGM}{low-density generator-matrix}
\acrodef{QLDPC}{quantum low-density parity-check}
\acrodef{VN}{variable node}
\acrodef{CN}{check node}
\acrodef{BP}{belief propagation}
\acrodef{CSS}{Calderbank–Shor–Steane}

\acrodef{GB}{generalized bicycle}
\acrodef{PCM}{parity check matrix}
\acrodefplural{PCM}[PCMs]{parity check matrices}
\acrodef{OSD}{ordered statistics decoding}
\acrodef{QEC}{quantum error correction}
\acrodef{QSC}{quantum stabilizer code}
\acrodef{CSS}{Calderbank–Shor–Steane}
\acrodef{NN}{neural network}
\acrodef{LLR}{log-likelihood ratio}
\acrodef{MWPM}{minimum weight perfect matching}
\acrodef{MS}{min-sum}
\acrodef{FG}{finite geometry}
\acrodef{EG}{Euclidean geometry}
\acrodef{PG}{projective geometry}

\acrodefplural{FG}[FGs]{finite geometries}
\acrodefplural{EG}[EGs]{Euclidean geometries}
\acrodefplural{PG}[PGs]{projective geometries}

\acrodef{URLLC}{ultra reliable and low latency communications}
\acrodef{QC}{quasi-cyclic}
\acrodef{CPM}{circulant permutation matrix}
\acrodef{ML}{maximum likelihood}

\acrodef{AWGN}{additive white Gaussian noise}
\acrodef{BMSC}{binary-input memoryless symmetric output channel}

\usepackage{mathtools}

%% file: figures/ISTC32_16.tex
\begin{tikzpicture}[scale=0.92,spy using outlines={rectangle, magnification=2}]

\begin{axis}[%
width=.8\columnwidth,
height=6cm,
at={(0.758in,0.645in)},
scale only axis,
xmin=1,
xmax=8,
xlabel style={font=\color{white!15!black}},
xlabel={$E_{\mathrm{b}}/N_0$ ($\si{dB}$)},
ymode=log,
ymin=1e-08,
ymax=1,
yminorticks=true,
ylabel style={font=\color{white!15!black}},
ylabel={FER},
axis background/.style={fill=white},
xmajorgrids,
ymajorgrids,
legend style={at={(0.03,0.03)}, anchor=south west, legend cell align=left, align=left, draw=white!15!black,font=\scriptsize}
]

\addplot [color=black,dashdotted, line width=1.1pt]
table[row sep=crcr]{%
0.00   3.125e-01\\  %
1.00   1.543e-01\\  %
2.00   6.983e-02\\  %
3.00   2.391e-02\\  %
4.00   6.332e-03\\  %
5.00   1.310e-03\\  %
6.00   2.647e-04\\  %
7.00   2.035e-05\\ %
8.00   1.341e-06\\
};
\addlegendentry{ML $\mathcal{C}_{\text{CCSDS}}$}

\addplot [color=KITblue,dashdotted,mark repeat=2, mark phase=2,line width=1.1pt,mark=triangle, mark options={solid, KITblue}]
table[row sep=crcr]{%
1.0  5.891016e-01\\
1.5  4.566210e-01\\
2.0  3.883495e-01\\
2.5  2.686367e-01\\
3.0  1.652893e-01\\
3.5  1.128668e-01\\
4.0  5.813109e-02\\
4.5  2.875216e-02\\
5.0  1.222868e-02\\
5.5  4.508007e-03\\
6.0  1.587339e-03\\
6.5  4.446910e-04\\
7.0  1.222306e-04\\
};
\addlegendentry{BP\--30 $\mathcal{C}_{\text{CCSDS}}$}

\addplot [color=black, line width=1.1pt]
table[row sep=crcr]{%
0	0.331858407079646\\
0.5	0.239043824701195\\
1	0.157397691500525\\
1.5	0.106496272630458\\
2	0.0591016548463357\\
2.5	0.03060599877576\\
3	0.0154966682163335\\
3.5	0.00715563506261181\\
4	0.00242977937603266\\
4.5	0.000928789694149554\\
5	0.00027793527443329\\
5.5	8.35736773003028e-05\\
6	2.31466103987999e-05\\
6.5	5.57120341522198e-06\\
7	1.4e-06\\
7.5  2.280000e-07\\
8.0  3.300000e-08\\
};
\addlegendentry{ML approx.}

\addplot [color=KITblue,mark=square, line width=1.1pt]
table[row sep=crcr]{
1.0  4.478649e-01\\
1.5  3.224044e-01\\
2.0  2.232916e-01\\
2.5  1.563719e-01\\
3.0  9.615742e-02\\
3.5  4.819064e-02\\
4.0  2.442670e-02\\
4.5  9.866178e-03\\
5.0  3.533517e-03\\
5.5  1.214520e-03\\
6.0  3.771289e-04\\
6.5  8.980000e-05\\
7.0  1.830000e-05\\
7.5  4.233333e-06\\
8.0  7.170000e-07\\
};\addlegendentry{BP-30}

\addplot [color=KITgreen!60,mark=asterisk,line width=1.1pt]
table[row sep=crcr]{
1.0  5.858987e-01\\
1.5  4.733607e-01\\
2.0  3.984601e-01\\
2.5  2.935631e-01\\
3.0  1.988120e-01\\
3.5  1.302736e-01\\
4.0  7.650553e-02\\
4.5  4.131367e-02\\
5.0  2.000544e-02\\
5.5  8.967366e-03\\
6.0  3.601028e-03\\
6.5  1.290000e-03\\
7.0  4.170000e-04\\
7.5  1.328941e-04\\
8.0  3.860900e-05\\
};\addlegendentry{GAED-3\--BP\--30   Path-2}

\addplot [color=KITgreen!40,mark=asterisk, line width=1.1pt]
table[row sep=crcr]{
1.0  6.156902e-01\\
1.5  5.068306e-01\\
2.0  4.138595e-01\\
2.5  3.221717e-01\\
3.0  2.231298e-01\\
3.5  1.452780e-01\\
4.0  9.200330e-02\\
4.5  4.984617e-02\\
5.0  2.518178e-02\\
5.5  1.177010e-02\\
6.0  4.939283e-03\\
6.5  1.906200e-03\\
7.0  6.630000e-04\\
7.5  2.153456e-04\\
8.0  6.254700e-05\\
};\addlegendentry{GAED-3\--BP\--30   Path-3}

\addplot [color=KITgreen, mark=asterisk,line width=1.1pt]
table[row sep=crcr]{%
1.0  3.972195e-01\\
1.5  2.732240e-01\\
2.0  1.924928e-01\\
2.5  1.300390e-01\\
3.0  7.425283e-02\\
3.5  3.530450e-02\\
4.0  1.573378e-02\\
4.5  6.062168e-03\\
5.0  1.944164e-03\\
5.5  5.577587e-04\\
6.0  1.381424e-04\\
6.5  2.670000e-05\\
7.0  4.086098e-06\\
7.5  5.836945e-07\\
 8.0  6.554450e-08\\
};
\addlegendentry{GAED-3\--BP\--30 }

\addplot [color=KITred!60,mark=diamond,mark repeat=2, mark phase=1, line width=1.1pt]
table[row sep=crcr]{%
1.0  5.135641e-01\\
1.5  4.147038e-01\\
2.0  3.030014e-01\\
2.5  2.199519e-01\\
3.0  1.426867e-01\\
3.5  8.813437e-02\\
4.0  4.605458e-02\\
4.5  2.239809e-02\\
5.0  8.424937e-03\\
5.5  3.284441e-03\\
6.0  1.103000e-03\\
6.5  3.355000e-04\\
7.0  9.220000e-05\\
7.5  2.321667e-05\\
8.0  5.872818e-06\\
};
\addlegendentry{iGAED-3\--BP\--30   Path-2}

\addplot [color=KITred!40, mark=diamond,mark repeat=2, mark phase=1,line width=1.1pt]
table[row sep=crcr]{%
1.0  5.322732e-01\\
1.5  4.475375e-01\\
2.0  3.272987e-01\\
2.5  2.394832e-01\\
3.0  1.616093e-01\\
3.5  1.000542e-01\\
4.0  5.398991e-02\\
4.5  2.574477e-02\\
5.0  1.124670e-02\\
5.5  4.284053e-03\\
6.0  1.553000e-03\\
6.5  4.708000e-04\\
7.0  1.359000e-04\\
7.5  3.415000e-05\\
8.0  8.337295e-06\\
};
\addlegendentry{iGAED-3\--BP\--30   Path-3}

\addplot [color=KITred, mark=diamond,line width=1.1pt]
table[row sep=crcr]{%
1.0  3.741815e-01\\
1.5  2.855103e-01\\
2.0  1.905669e-01\\
2.5  1.201923e-01\\
3.0  6.818957e-02\\
3.5  3.612064e-02\\
4.0  1.483239e-02\\
4.5  5.486366e-03\\
5.0  1.614746e-03\\
5.5  4.251092e-04\\
6.0  8.974702e-05\\
6.5  1.541916e-05\\
7.0  2.631018e-06\\
7.5  3.607073e-07\\
 8.0  4.880078e-08\\
};
\addlegendentry{iGAED-3\--BP\--30 }

\coordinate (spypoint) at (axis cs:4.9,0.0005);
    \coordinate (spyviewer) at (axis cs:6,0.02);	
    \spy[width=1.7cm,height=1.25cm, thin, spy connection path={\draw(tikzspyonnode.south west) -- (tikzspyinnode.south west);\draw (tikzspyonnode.south east) -- (tikzspyinnode.south east);
    \draw (tikzspyonnode.north west) -- (tikzspyinnode.north west);\draw (tikzspyonnode.north east) -- (intersection of  tikzspyinnode.north east--tikzspyonnode.north east and tikzspyinnode.south east--tikzspyinnode.south west);
    ;}] on (spypoint) in node at (spyviewer);
\coordinate (a) at ($(axis cs:-10.8/1.4,-0.12)+(spyviewer)$);
\coordinate[label={[font=\small,text=black]right:$10^{-2}$}] (b) at ($(axis cs:+10.8/1.4,-0.12)+(spyviewer)$);

\end{axis}
\end{tikzpicture}%

%% file: figures/ISTC63_45.tex
\begin{tikzpicture}[scale=0.92]

\begin{axis}[%
width=.8\columnwidth,
height=6cm,
at={(0.758in,0.645in)},
scale only axis,
xmin=2,
xmax=7,
xlabel style={font=\color{white!15!black}},
xlabel={$E_{\mathrm{b}}/N_0$ ($\si{dB}$)},
ymode=log,
ymin=1e-07,
ymax=1,
yminorticks=true,
ylabel style={font=\color{white!15!black}},
ylabel={FER},
axis background/.style={fill=white},
xmajorgrids,
ymajorgrids,
legend style={at={(0.03,0.03)}, anchor=south west, legend cell align=left, align=left, draw=white!15!black,font=\scriptsize}
]

\addplot [color=black, dashdotted,line width=1pt]
table[row sep=crcr]{%
0.00  6.329e-01\\
0.50  4.975e-01\\
1.00  3.704e-01\\
1.50  2.445e-01\\
2.00  1.447e-01\\
2.50  7.353e-02\\
3.00  2.595e-02\\
3.50  7.918e-03\\
4.00  2.134e-03\\
4.50  4.751e-04\\
5.00  5.337e-05\\
5.50  6.300e-06\\
};
\addlegendentry{ML $\mathcal{C}_{\mathrm{BCH}}$}

\addplot [color=KITblue,dashdotted,mark repeat=2, mark phase=2, line width=1pt,mark=triangle, mark options={solid,KITblue}]
table[row sep=crcr]{%
0.0 9.259259259259259300e-01\\
0.5 9.287925696594426794e-01\\
1.0 8.219178082191780366e-01\\
1.5 6.564551422319474527e-01\\
2.0 5.244755244755244794e-01\\
2.5 3.640776699029126262e-01\\
3.0 2.427184466019417508e-01\\
3.5 1.220504475183075699e-01\\
4.0 6.384337092998509933e-02\\
4.5 2.570253598355037861e-02\\
5.0 7.212231945379363705e-03\\
5.5 2.290006412017953855e-03\\
6.0 4.538124786329958112e-04\\
6.5 8.600000000000000331e-05\\
7.0 2.000000000000000164e-05\\
};
\addlegendentry{BP\--30 $\mathcal{C}_{\mathrm{BCH}}$}

\addplot [color=black, line width=1pt]
table[row sep=crcr]{%
0	6.91358025e-01\\
0.5 5.67951318e-01\\
1   4.28790199e-01\\
1.5 2.59981430e-01\\
2   1.61290323e-01\\
2.5 6.74698795e-02\\
3   2.75834893e-02\\
3.5 9.06618314e-03\\
4   2.19662974e-03\\
4.5 5.14540359e-04\\
5   9.50628433e-05\\
5.5 1.72278718e-05\\
6   2.31674393e-06\\ 
};
\addlegendentry{ML approx.}

\addplot [color=KITblue,mark=square,mark repeat=2, mark phase=1, line width=1pt]
table[row sep=crcr]{
0.0 9.404388714733542542e-01\\
0.5 9.036144578313253239e-01\\
1.0  7.956204e-01\\
1.5  6.848875e-01\\
2.0  5.218391e-01\\
2.5  3.889764e-01\\
3.0  2.609626e-01\\
3.5  1.382532e-01\\
4.0  6.456756e-02\\
4.5  2.308941e-02\\
5.0  8.219097e-03\\
5.5  2.222883e-03\\
6.0  5.207266e-04\\
6.5  1.185499e-04\\
7.0  2.265000e-05\\
};\addlegendentry{BP-30}

\addplot [color=KITgreen!60,mark=asterisk, line width=1pt]
table[row sep=crcr]{
1.0  8.029197e-01\\
1.5  7.202572e-01\\
2.0  5.839080e-01\\
2.5  4.472441e-01\\
3.0  3.336898e-01\\
3.5  1.751518e-01\\
4.0  8.648864e-02\\
4.5  3.944441e-02\\
5.0  1.434633e-02\\
5.5  4.773833e-03\\
6.0  1.458714e-03\\
6.5  4.274139e-04\\
7.0  1.252500e-04\\
};\addlegendentry{GAED-3\--BP-30 Path-2}

\addplot [color=KITgreen!40, mark=asterisk,line width=1pt]
table[row sep=crcr]{
1.0  7.956204e-01\\
1.5  7.459807e-01\\
2.0  6.068966e-01\\
2.5  4.582677e-01\\
3.0  3.187166e-01\\
3.5  1.854274e-01\\
4.0  8.768434e-02\\
4.5  3.836209e-02\\
5.0  1.553320e-02\\
5.5  5.095768e-03\\
6.0  1.525000e-03\\
6.5  4.310956e-04\\
7.0  1.125500e-04\\
};\addlegendentry{GAED-3\--BP-30  Path-3}

\addplot [color=KITgreen, mark=asterisk,line width=1pt]
table[row sep=crcr]{%
1.0  7.380074e-01\\
1.5  6.379585e-01\\
2.0  4.914005e-01\\
2.5  3.110420e-01\\
3.0  1.851852e-01\\
3.5  8.859358e-02\\
4.0  3.734479e-02\\
4.5  1.113741e-02\\
5.0  3.044858e-03\\
5.5  6.075473e-04\\
 6.0  8.940401e-05\\ %
 6.5  1.035825e-05\\
 7.0  1.034451e-06\\
};
\addlegendentry{GAED-3\--BP-30}

\addplot [color=KITred!60,mark=diamond,mark repeat=2, mark phase=1, line width=1pt]
table[row sep=crcr]{%
1.0  7.845745e-01\\
1.5  6.604555e-01\\
2.0  5.392157e-01\\
2.5  4.143378e-01\\
3.0  2.796610e-01\\
3.5  1.458489e-01\\
4.0  7.458340e-02\\
4.5  2.836642e-02\\
5.0  1.119776e-02\\
5.5  3.021136e-03\\
6.0  8.027773e-04\\
6.5  1.490843e-04\\
7.0  2.951829e-05\\
};
\addlegendentry{iGAED-3\--BP\--30 Path-2}

\addplot [color=KITred!40,mark=diamond,mark repeat=2, mark phase=1, line width=1pt]
table[row sep=crcr]{%
1.0  7.606383e-01\\
1.5  6.356108e-01\\
2.0  5.245098e-01\\
2.5  3.961118e-01\\
3.0  3.050847e-01\\
3.5  1.780105e-01\\
4.0  8.412878e-02\\
4.5  3.257594e-02\\
5.0  9.833069e-03\\
5.5  2.815326e-03\\
6.0  7.572320e-04\\
6.5  1.552318e-04\\
7.0  2.995438e-05\\
};
\addlegendentry{iGAED-3\--BP\--30 Path-3}

\addplot [color=KITred, mark=diamond,line width=1pt]
table[row sep=crcr]{%
 1.0  5.925926e-01\\
 1.5  4.371585e-01\\
 2.0  3.152088e-01\\
 2.5  2.347418e-01\\
 3.0  1.526718e-01\\
 3.5  8.178287e-02\\
 4.0  3.323916e-02\\
 4.5  1.146197e-02\\
 5.0  2.873873e-03\\
 5.5  5.195898e-04\\
 6.0  8.092464e-05\\
6.5  8.840283e-06\\
 7.0  6.602978e-07\\
};
\addlegendentry{iGAED-3\--BP\--30 }

\end{axis}
\end{tikzpicture}%

%% file: figures/ISTC32_16_diff_paths.tex
\begin{tikzpicture}[scale=0.92]

\begin{axis}[%
width=.8\columnwidth,
height=6cm,
at={(0.758in,0.645in)},
scale only axis,
xmin=4,
xmax=7,
xlabel style={font=\color{white!15!black}},
xlabel={$E_{\mathrm{b}}/N_0$ ($\si{dB}$)},
ymode=log,
ymin=1e-06,
ymax=1e-1,
yminorticks=true,
ylabel style={font=\color{white!15!black}},
ylabel={FER},
axis background/.style={fill=white},
xmajorgrids,
ymajorgrids,
legend style={at={(0.03,0.03)}, anchor=south west, legend cell align=left, align=left, draw=white!15!black,font=\scriptsize}
]
\addplot [color=black, line width=1.1pt]
table[row sep=crcr]{%
0	0.331858407079646\\
0.5	0.239043824701195\\
1	0.157397691500525\\
1.5	0.106496272630458\\
2	0.0591016548463357\\
2.5	0.03060599877576\\
3	0.0154966682163335\\
3.5	0.00715563506261181\\
4	0.00242977937603266\\
4.5	0.000928789694149554\\
5	0.00027793527443329\\
5.5	8.35736773003028e-05\\
6	2.31466103987999e-05\\
6.5	5.57120341522198e-06\\
7	1.4e-06\\
7.5	2.3e-07\\
8	0\\
};
\addlegendentry{ML}

\addplot [color=KITgreen,mark=asterisk, line width=1.1pt]
table[row sep=crcr]{%
1.0  3.972195e-01\\
1.5  2.732240e-01\\
2.0  1.924928e-01\\
2.5  1.300390e-01\\
3.0  7.425283e-02\\
3.5  3.530450e-02\\
4.0  1.573378e-02\\
4.5  6.062168e-03\\
5.0  1.944164e-03\\
5.5  5.577587e-04\\
6.0  1.381424e-04\\
6.5  2.670000e-05\\
7.0  4.086098e-06\\
7.5  8.000000e-07\\ %
};
\addlegendentry{GAED-3\--BP\--30 }

\addplot [color=KITgreen,dashed,mark=asterisk,  line width=1.1pt]
table[row sep=crcr]{%
2.0  1.754386e-01\\
2.5  1.158749e-01\\
3.0  6.281407e-02\\
3.5  3.016591e-02\\
4.0  1.466544e-02\\
4.5  4.976858e-03\\
5.0  1.814339e-03\\
5.5  4.818818e-04\\
6.0  1.089934e-04\\
6.5  2.170590e-05\\
7.0  3.043346e-06\\
};
\addlegendentry{GAED-5\--BP\--30   }

\addplot [color=KITred, mark=diamond,line width=1.1pt]
table[row sep=crcr]{%
1.0  3.741815e-01\\
1.5  2.855103e-01\\
2.0  1.905669e-01\\
2.5  1.201923e-01\\
3.0  6.818957e-02\\
3.5  3.612064e-02\\
4.0  1.483239e-02\\
4.5  5.486366e-03\\
5.0  1.614746e-03\\
5.5  4.251092e-04\\
6.0  8.974702e-05\\
6.5  1.541916e-05\\
7.0  2.631018e-06\\
7.5  3.000000e-07\\
};
\addlegendentry{iGAED-3\--BP\--30 }

\addplot [color=KITred,dashed, mark=diamond,line width=1.1pt]
table[row sep=crcr]{%
2.0  1.002903e-01\\
2.5  7.239474e-02\\
3.0  5.070047e-02\\
3.5  3.127829e-02\\
4.0  1.323765e-02\\
4.5  4.705824e-03\\
5.0  1.303008e-03\\
5.5  3.923148e-04\\
6.0  7.638556e-05\\
6.5  1.263589e-05\\
7.0  2.237034e-06\\
};
\addlegendentry{iGAED-5\--BP\--30 }

\end{axis}
\end{tikzpicture}%

%% file: figures/ISTC63_45_diff_iter.tex
\begin{tikzpicture}[scale=0.92]

\begin{axis}[%
width=.8\columnwidth,
height=6cm,
at={(0.758in,0.645in)},
scale only axis,
xmin=5,
xmax=7,
xlabel style={font=\color{white!15!black}},
xlabel={$E_{\mathrm{b}}/N_0$ ($\si{dB}$)},
ymode=log,
ymin=1e-06,
ymax=1e-2,
yminorticks=true,
ylabel style={font=\color{white!15!black}},
ylabel={FER},
axis background/.style={fill=white},
xmajorgrids,
ymajorgrids,
legend style={at={(0.03,0.03)}, anchor=south west, legend cell align=left, align=left, draw=white!15!black,font=\scriptsize}
]

\addplot [color=KITblue,mark=square, line width=1pt]
table[row sep=crcr]{
0.0 9.404388714733542542e-01\\
0.5 9.036144578313253239e-01\\
1.0  7.956204e-01\\
1.5  6.848875e-01\\
2.0  5.218391e-01\\
2.5  3.889764e-01\\
3.0  2.609626e-01\\
3.5  1.382532e-01\\
4.0  6.456756e-02\\
4.5  2.308941e-02\\
5.0  8.219097e-03\\
5.5  2.222883e-03\\
6.0  5.207266e-04\\
6.5  1.185499e-04\\
7.0  2.265000e-05\\
};\addlegendentry{BP-30}

\addplot [color=KITgreen,mark=asterisk, dashed, line width=1.1pt]
table[row sep=crcr]{%
3.0  2.466091e-01\\
3.5  1.387925e-01\\
4.0  6.148171e-02\\
4.5  2.073613e-02\\
5.0  5.956813e-03\\
5.5  1.416190e-03\\
6.0  2.253260e-04\\
6.5  3.149035e-05\\
7.0  3.513091e-06\\
};
\addlegendentry{GAED-5\--BP\--6   }

\addplot [color=KITred,mark=diamond,dashed, line width=1.1pt]
table[row sep=crcr]{%
4.0  5.474952e-02\\
4.5  2.134927e-02\\
5.0  5.696383e-03\\
5.5  1.202516e-03\\
6.0  2.344719e-04\\
6.5  2.921687e-05\\
7.0  2.905630e-06\\
};
\addlegendentry{iGAED-5\--BP\--6   }

\end{axis}
\end{tikzpicture}%